\documentclass[preprint,12pt]{elsarticle}

\usepackage{amsmath,xcolor}
\usepackage{amssymb}

\usepackage{color,soul}

\usepackage{lmodern}

\journal{Physics Letters A}

\begin{document}

\begin{frontmatter}

\title{Pauli equation in spaces of constant curvature and extended  Nikiforov-Uvarov method} %% Article title

\author[a,b,c]{Abdaljalel E. Alizzi}
\ead{abdaljalel90@gmail.com}
\affiliation[a]{organization={Budker Institute of Nuclear Physics}, city={Novosibirsk}, postcode={630 090}, country={Russia}}
\affiliation[b]{organization={Department of Physics, Novosibirsk State University}, city={Novosibirsk},postcode={630 090}, country={Russia}}
\affiliation[c]{organization={Department of Physics, Al Furat University}, city={Deir-ez-Zor}, country={Syrian Arab Republic}}
\author[a,b]{Zurab K. Silagadze}
\ead{Z.K.Silagadze@inp.nsk.su}

\begin{abstract}
We apply the extended Nikiforov–Uvarov method to the non-relativistic limit of the Dirac equation with a Coulomb potential in spaces of constant curvature. In this case, the radial equation reduces to the Heun equation, and the extended Nikiforov–Uvarov method easily yields a quantization condition which leads to necessary condition under which the resulting Heun equation can have polynomial solutions. The energy spectrum implied by the quantization condition is virtually identical to the spectrum of a spinless particle obtained using the Schr\"{o}dinger equation, except for the absence of the ``geometric potential'', confirming the non-commutativity of the naive non-relativistic limit with the ``squaring'' of the Dirac equation, first discovered on curved surfaces. However, the necessary conditions for the existence of polynomial solutions cannot be met, and this fact undermines the reliability of the results obtained. This circumstance forces us to conclude that the extended Nikiforov–Uvarov method has limited, if any, value when considering similar problems in quantum mechanics.
\end{abstract}


%\item Pauli particle, bound by the Coulomb potential in spaces of constant curvature, is considered;
%\item Application of quantization condition of the extended Nikiforov–Uvarov method leads to the reasonable energy spectrum;
%\item Unfortunately, the obtained results cannot be strictly justified, which undermines the usefulness of the extended Nikiforov–Uvarov method.

%% Keywords
\begin{keyword}
Extended Nikiforov–Uvarov method; Pauli equation in spaces of constant curvature; Heun equation. 



\end{keyword}

\end{frontmatter}



\section{Introduction}
The Nikiforov-Uvarov method \cite{Nikiforov_1988} is a popular and useful tool for the analysis of second-order differential equations of generalized hypergeometric type in some problems of quantum \cite{Ellis_2023,Berkdemir_2012,Suslov_2020,Tezcan_2009,Miranda_2010,Zhang_2010} and classical \cite{Zhang_2025} physics. In the previous publication \cite{Alizzi_2025} we applied the Nikiforov-Uvarov method to a non-relativistic spinless Schr\"{o}dinger particle bound by a Coulomb potential in spaces of constant curvature. A natural next step is to include spin and consider the non-relativistic Pauli equation in the same situation. However, in this case, we obtain a Heun equation with four singularities \cite{Ovsiyuk_2011}, and the Nikiforov-Uvarov method in its usual form is inapplicable.

Extension of Nikiforov-Uvarov method for the solution of Heun equation was considered in \cite{Karayer_2015}. The aim of this paper is to apply this extended Nikiforov-Uvarov method to the non-relativistic limit of the Dirac equation (i.e., to the Pauli equation) with Coulomb potential in spaces of constant curvature. This problem was considered by a different method in \cite{Ovsiyuk_2011}.  We show that the extended Nikiforov-Uvarov method reproduces the results of \cite{Ovsiyuk_2011}, thus appearing to allay the concerns of \cite{Fernandez_2021,Fernandez_2026} about the uselessness of this method for solving eigenvalue problems in quantum mechanics. However, the results obtained in this way cannot be mathematically rigorously justified, and we come to agree with the main conclusion of \cite{Fernandez_2026} that the extended Nikiforov-Uvarov method, although it may have some heuristic value, is in a strict sense useless in problems of quantum mechanics leading to equations of the Heun type.

\section{Dirac equation in spaces of constant curvature}
As in \cite{Alizzi_2025}, we use generalized trigonometric functions to express the line element in space-times with constant curvature spaces in the form
\begin{equation}
ds^2=dt^2-dr^2-S_{\kappa}^2(r)\left (d\theta^2+\sin^2{\theta}\,d\varphi^2\right ).
\label{eq1}
\end{equation}
The generally covariant Dirac equation in a curved space-time  has the form \cite{parker_2009}
\begin{equation}
\left[i \, e_a^{\;\;\mu} \;\gamma^{a} \left ( \nabla_{\mu}+ieA_\mu\right)-m\right] \Psi = 0. 
    \label{eq2}
\end{equation}
Here  $A_\mu$ represents the electromagnetic four-vector, and the covariant derivative of the Dirac spinor field $\Psi(x)$ is equal to
\begin{equation}
    \nabla_{\mu} \Psi = \partial _{\mu} \Psi + i \;\Omega_{\mu a b} \; \Sigma^{ab} \,\Psi. \label{eq3}
\end{equation}
We will use Latin indices for the tensor components in the local orthonormal frame, in which the metric tensor is
\begin{equation}
    \eta_{ab} = \mathrm{diag}(1,-1,-1,-1), 
    \label{eq4}
\end{equation}
while, according to (\ref{eq1}), in the original coordinates $(t,r,\theta,\varphi)$ the components of the metric and its inverse are
\begin{eqnarray} &&
g_{\mu\nu}=\mathrm{diag}\left(1,-1,-S_\kappa(r)^2,-S_\kappa(r)^2\sin^2{\theta}\right ),
\nonumber \\ &&
g^{\mu\nu}=\mathrm{diag}\left(1,-1,-\frac{1}{S_\kappa(r)^2},-\frac{1}{S_\kappa(r)^2\sin^2{\theta}}\right ).
\label{eq5}
\end{eqnarray}
Other symbols in (\ref{eq3}) are defined as follows.
\begin{equation}
    \Sigma^{ab} = - \dfrac{i}{8} \; \left[\gamma^{a} , \gamma^{b} \right],
    \label{eq6}
\end{equation}
and the Christoffel symbols and spin-connection coefficients are given by the usual expressions 
\begin{eqnarray} &&
  \Gamma^{\alpha}_{\; \mu \nu } = \frac{1}{2}\, g^{\alpha \beta} \left( \partial_{\mu} g_{\beta \nu} + \partial_{\nu} g_{\beta \mu} - \partial_{\beta} g_{\mu \nu }\right), \nonumber\\ &&
   \Omega_{\mu \; b}^{\;a} = e_{b}^{\; \rho} \; e^{a}_{\; \nu} \; \Gamma^{\nu}_{\; \mu \rho} - e_{b}^{\; \nu} \; \partial_{\mu} e^{a}_{\; \nu}=-e_{b}^{\; \nu}e^{a}_{\; \nu\,;\,\mu} .
\label{eq7}
\end{eqnarray}
The vierbein $e_{a} ^{\; \mu}$ satisfies $e_{a} ^{\,\, \mu} g_{\mu \nu} e_{b} ^{\,\, \nu} = \eta_{ab}$ and is defined up to a Lorentz transformation in the local inertial frame. In the case of metric (\ref{eq5}), the vierbein can be chosen in the diagonal form
\begin{eqnarray} &&
e_0^{\;\mu}=(1,0,0,0),\;e_1^{\;\mu}=(0,1,0,0), \nonumber \\ &&
e_2^{\;\mu}=\left (0,0,\frac{1}{S_\kappa(r)},0\right ),\,
e_3^{\;\mu}=\left (0,0,0,\frac{1}{S_\kappa(r)\sin{\theta}}\right ),
\label{eq8}
\end{eqnarray}
and their inverses are
\begin{eqnarray}  &&
e_{\;0}^{\mu}=(1,0,0,0),\;e_{\;1}^{\mu}=(0,1,0,0), \nonumber \\ &&
e_{\;2}^{\mu}=\left (0,0,S_\kappa(r),0\right ),\,
e_{\;3}^{\mu}=\left (0,0,0,S_\kappa(r)\sin{\theta}\right ),
\label{eq9}
\end{eqnarray}
As usual, constant gamma matrices in the local inertial frame satisfy the Clifford algebra 
\begin{equation}
\gamma^{a} \gamma^{b} + \gamma^{b} \gamma^{a} = 2 \, \eta^{ab}.
\label{eq10}
\end{equation}
Although we will not need them, we can compute the non-zero Christoffel symbols (we only list them for completeness and comparison with \cite{Bessis_1982})
\begin{eqnarray} &&
\Gamma_{\;r\theta}^\theta=\Gamma_{\;\theta r}^\theta=\Gamma_{\;r\varphi}^\varphi=\Gamma_{\;\varphi r}^\varphi=\frac{1}{T_\kappa(r)},\;\; \Gamma_{\;\theta\varphi}^\varphi=\Gamma_{\;\varphi \theta}^\varphi=\frac{1}{\tan{\theta}}, \\ &&
\Gamma_{\;\theta\theta}^r=-S_\kappa(r)C_\kappa(r),\;\;\Gamma_{\;\varphi\varphi}^r=-S_\kappa(r)C_\kappa(r)\sin^2{\theta},\;\;\Gamma_{\;\varphi\varphi}^\theta=-\sin{\theta}\cos{\theta}.
\nonumber
\label{eq11}
\end{eqnarray}
If we introduce notations (note that, since $g_{\nu\rho\,;\,\mu}=0$, $g_{\nu\rho}e_{a\;;\mu}^{\;\rho}=(g_{\nu\rho}e_{a}^{\;\rho})_{;\mu}=e_{a\nu\,;\,\mu}$)
\begin{equation}
\gamma^{ab}=\frac{1}{4}[\gamma^a,\gamma^b],\;\;\;\gamma_{abc}=\Omega_{\mu a b}\,e_c^{\;\mu}=-e_{a\nu\,;\,\mu}e_b^{\;\nu}e_c^{\;\mu},  
\label{eq12} 
\end{equation}
then the Dirac equation (\ref{eq2}) takes the form \cite{Ovsiyuk_2011}
\begin{equation}
\left \{ i\gamma^c\left [e_c^{\;\mu}(\partial_\mu+ieA_\mu)+\frac{1}{2}\gamma_{abc}\gamma^{ab}\right ]-m \right \}\Psi =0.
\label{eq13}
\end{equation}
Since $0=(\eta_{ab})_{;\mu}=(e_{a\nu}e_b^{\;\nu})_{;\mu}=e_{a\nu\,;\,\mu}e_b^{\;\nu}+e_{b\nu\,;\,\mu}e_a^{\;\nu}$, the Ricci rotation coefficients $\gamma_{abc}$ are antisymmetric in the first two indices: $\gamma_{abc}=-\gamma_{bac}$. Any object with this property satisfies
\begin{equation}
\gamma_{abc}=\frac{1}{2}\left ( \,\gamma_{a[bc]}-\gamma_{c[ab]}+ \gamma_{b[ca]}\, \right ).
\label{eq14}
\end{equation}
The importance of (\ref{eq14}) is that it allows one to calculate Ricci rotation coefficients without using Christoffel symbols \cite{Chandrasekhar_1983}. Indeed, since $\Gamma^{\nu}_{\; \mu \rho}=\Gamma^{\nu}_{\; \rho \mu}$, it can be easily checked that
\begin{equation}
 \gamma_{a[bc]}=   \lambda_{a[bc]},\;\;\;\lambda_{abc}=e_{a\nu\,,\,\mu}e_b^{\;\nu}e_c^{\;\mu},
 \label{eq15}
\end{equation}
that is, $\lambda_{abc}$ contains the ordinary derivative, not the covariant derivative. Using (\ref{eq8}) and (\ref{eq9}) , we find that the only non-zero $\lambda_{abc}$ coefficients are
\begin{equation}
\lambda_{221}=\lambda_{331}=\frac{1}{T_\kappa(r)},\;\;\; \lambda_{332}=\frac{1}{S\kappa(r)\tan{\theta}}.
\label{eq16}
\end{equation}
Then (\ref{eq14}) and (\ref{eq15}) lead to the following non-zero Ricci rotation coefficients
\begin{equation}
\gamma_{122}=-\gamma_{212}=\gamma_{133}=-\gamma_{313}=\frac{1}{T_\kappa(r)},\;\;
\gamma_{233}=-\gamma_{323}=\frac{1}{S\kappa(r)\tan{\theta}}.
\label{eq17}
\end{equation}
Therefore, the Dirac equation (\ref{eq13}) takes the form (we assume $\vec{A}=0)$
\begin{equation}
\left \{ i\gamma^0\left (\frac{\partial}{\partial t}+ieA_0\right)+i\gamma^1\left(\frac{\partial}{\partial r}+\frac{1}{T_\kappa(r)}\right )+\frac{\hat \Sigma}{S_\kappa(r)}-m \right \}\Psi =0,
\label{eq18}
\end{equation}
where the angular operator $\hat\Sigma$ is related to the Brill-Wheeler angular momentum operator \cite{Brill_1957,Villalba_1994} and has the form
\begin{equation}
\hat{\Sigma}=i\gamma^2\frac{\partial}{\partial \theta}+\gamma^3\frac{i\frac{\partial}{\partial \varphi}+i\cos{\theta}\,\gamma^{23}}{\sin{\theta}}.
\label{eq19}
\end{equation}
This equation can be somewhat simplified if we assume
\begin{equation}
\Psi(t,r,\theta,\varphi)=\frac{1}{S_\kappa(r)}\,\bar\Psi(t,r,\theta,\varphi),
\label{eq20}
\end{equation}
so that
\begin{equation}
\left(\frac{\partial}{\partial r}+\frac{1}{T_\kappa(r)}\right )\Psi= \frac{1}{S_\kappa(r)}\,\frac{\partial \bar\Psi}{\partial r}.
\label{eq21}
\end{equation}
As a result, (\ref{eq18}) takes the form
\begin{equation}
 \left \{ i\gamma^0\left (\frac{\partial}{\partial t}+ieA_0\right)+i\gamma^1\frac{\partial}{\partial r}+\frac{\hat \Sigma}{S_\kappa(r)}-m \right \}\bar \Psi =0.
\label{eq22}
\end{equation}   

\section{Separation of variables and radial equations}
In the gamma matrix representation, where $\gamma^{23}$ in (\ref{eq19}) has the diagonal form $\gamma^{23}=\frac{1}{2}\sigma_3\oplus\sigma_3$, it can be shown that the most general spinor $\Psi$ with a definite total momentum $j$ and its third component $m$ corresponds to \cite{Redkov_1998,Redkov_2010}
\begin{equation}
\bar\Psi_{jm}(t,r,\theta,\varphi)=\left (\begin{array}{c} f_1(t,r)\,D_{-m,-1/2}^j(\varphi,\theta,0) \\
f_2(t,r)\,D_{-m,\;1/2}^j(\varphi,\theta,0) \\
f_3(t,r)\,D_{-m,-1/2}^j(\varphi,\theta,0) \\
f_4(t,r)\,D_{-m,\;1/2}^j(\varphi,\theta,0) \\
 \end{array}\right ),
 \label{eq23}
\end{equation}
where $D^j_{mm^\prime}(\varphi,\theta,\psi)$ are Wigner D-functions \cite{Varshalovich_1988}. The needed representation of gamma matrices (which can be called Schr\"{o}dinger representation \cite{Schrodinger_1938,Redkov_1998}) can be obtained from the usual chiral (Weyl) representation $\bar{\gamma}^k$ by the similarity transformation \cite{Villalba_1994}
\begin{equation}
 \gamma^0=\bar{\gamma}^0=\left (\begin{array}{cc} 0 & 1 \\ 1 & 0 \end{array}\right ),\;\gamma^k=S\bar{\gamma}^kS^{-1},\;S=\frac{1}{2}\left(1-\bar{\gamma}^1 \bar{\gamma}^2+\bar{\gamma}^1 \bar{\gamma}^3-\bar{\gamma}^2 \bar{\gamma}^3\right).
 \label{eq24}
\end{equation}
It can be checked that $\bar{\gamma}^3S\bar{\gamma}^1=\bar{\gamma}^1S\bar{\gamma}^2=\bar{\gamma}^2S\bar{\gamma}^3=-S$, which implies $\gamma^1=\bar{\gamma}^3$, $\gamma^2=\bar{\gamma}^1$, $\gamma^3=\bar{\gamma}^2$, or
\begin{equation}
\gamma^1=\left (\begin{array}{cc} 0 & -\sigma_3 \\ \sigma_3 & 0 \end{array}\right ),\;\;\gamma^2=\left (\begin{array}{cc} 0 & -\sigma_1 \\ \sigma_1 & 0 \end{array}\right ),\;\;\gamma^3=\left (\begin{array}{cc} 0 & -\sigma_2 \\ \sigma_2 & 0 \end{array}\right ).
\label{eq25}
\end{equation}
Alternatively, we can keep the gamma matrices in the Weyl representation, but start with the rotated vierbein fields $\bar e_0^\mu=e_0^\mu$, $\bar e_1^\mu=e_2^\mu$, $\bar e_2^\mu=e_3^\mu$, and $\bar e_3^\mu=e_1^\mu$ \cite{Redkov_1998,Ovsiyuk_2011}.

Using gamma matrices from (\ref{eq25}), $i\frac{\partial}{\partial \varphi}D^j_{-m,\,\pm 1/2}=-m\,D^j_{-m,\,\pm 1/2}$,
and the following recursive relations \cite{Varshalovich_1988}
\begin{eqnarray} &&
\frac{m+\frac{1}{2}\cos{\theta}}{\sin{\theta}}D^j_{-m,\,\frac{1}{2}}=\frac{1}{2}\left(j+\frac{1}{2}\right )D^j_{-m,\,-\frac{1}{2}}+\frac{1}{2}\sqrt{j^2+j-\frac{3}{4}}\,D^j_{-m,\,\frac{3}{2}}, \nonumber \\ &&
\frac{-m+\frac{1}{2}\cos{\theta}}{\sin{\theta}}D^j_{-m,\,-\frac{1}{2}}=-\frac{1}{2}\left(j+\frac{1}{2}\right )D^j_{-m,\,\frac{1}{2}}-\frac{1}{2}\sqrt{j^2+j-\frac{3}{4}}\,D^j_{-m,\,-\frac{3}{2}},\nonumber \\ &&
\frac{\partial}{\partial \theta}\,D^j_{-m,\,\frac{1}{2}}=\frac{1}{2}\left(j+\frac{1}{2}\right )D^j_{-m,\,-\frac{1}{2}}-\frac{1}{2}\sqrt{j^2+j-\frac{3}{4}}\,D^j_{-m,\,\frac{3}{2}}, \nonumber \\ &&
\frac{\partial}{\partial \theta}\,D^j_{-m,\,-\frac{1}{2}}=-\frac{1}{2}\left(j+\frac{1}{2}\right )D^j_{-m,\,\frac{1}{2}}+\frac{1}{2}\sqrt{j^2+j-\frac{3}{4}}\,D^j_{-m,\,-\frac{3}{2}},
\label{eq26}
\end{eqnarray}
we get
\begin{equation}
\hat \Sigma \left (\begin{array}{c} f_1\,D_{-m,-1/2}^j \\
f_2\,D_{-m,\;1/2}^j \\
f_3\,D_{-m,-1/2}^j \\
f_4\,D_{-m,\;1/2}^j\end{array}\right )= i\left(j+\frac{1}{2}\right)\left (\begin{array}{c} -f_4\,D_{-m,-1/2}^j \\
\;\;f_3\,D_{-m,\;1/2}^j \\
\;\;f_2\,D_{-m,-1/2}^j \\
-f_1\,D_{-m,\;1/2}^j\end{array}\right ).
\label{eq27}
\end{equation}
Then (\ref{eq22}) is reduced to the following system of radial equations \cite{Redkov_1998,Ovsiyuk_2011}
\begin{eqnarray} &&
i\left(\frac{\partial}{\partial t}+ieA_0\right)f_3-i\frac{\partial f_3}{\partial r}-i\left(j+\frac{1}{2}\right)\frac{f_4}{S_\kappa(r)}-mf_1=0,  \nonumber \\ &&
i\left(\frac{\partial}{\partial t}+ieA_0\right)f_4+i\frac{\partial f_4}{\partial r}+i\left(j+\frac{1}{2}\right)\frac{f_3}{S_\kappa(r)}-mf_2=0,  \nonumber \\ &&
i\left(\frac{\partial}{\partial t}+ieA_0\right)f_1+i\frac{\partial f_1}{\partial r}+i\left(j+\frac{1}{2}\right)\frac{f_2}{S_\kappa(r)}-mf_3=0,  \nonumber \\ &&
i\left(\frac{\partial}{\partial t}+ieA_0\right)f_2-i\frac{\partial f_2}{\partial r}-i\left(j+\frac{1}{2}\right)\frac{f_1}{S_\kappa(r)}-mf_4=0.
\label{eq28}
\end{eqnarray}
A further simplification can be achieved by requiring that solutions of equation (\ref{eq22}) have certain parity properties. The space inversion in spherical coordinates is given by transformation $\theta\to\theta^\prime=\pi-\theta$, $\varphi\to\varphi^\prime=\pi+\varphi$. Therefore, the Dirac operator 
\begin{equation}
\hat D=i\gamma^0\left (\frac{\partial}{\partial t}+ieA_0\right)+i\gamma^1\frac{\partial}{\partial r}+\frac{1}{S_\kappa(r)}\left [ i\gamma^2\frac{\partial}{\partial \theta}+\gamma^3\frac{i\frac{\partial}{\partial \varphi}+i\cos{\theta}\,\gamma^{23}}{\sin{\theta}} \right ]-m,
\label{eq29}
\end{equation}
is changed under this transformation to
\begin{equation}
\hat D^\prime=i\gamma^0\left (\frac{\partial}{\partial t}+ieA_0\right)+i\gamma^1\frac{\partial}{\partial r}+\frac{1}{S_\kappa(r)}\left [ -i\gamma^2\frac{\partial}{\partial \theta}+\gamma^3\frac{i\frac{\partial}{\partial \varphi}-i\cos{\theta}\,\gamma^{23}}{\sin{\theta}} \right ]-m.
\label{eq30}
\end{equation}
The transformed wave function $\bar \Psi^\prime(t,r,\theta^\prime,\varphi^\prime)$ must satisfy the same Dirac equation $\hat D^\prime \bar\Psi^\prime(t,r,\theta^\prime,\varphi^\prime)=0$ if we want parity to be conserved. Suppose we found such a unitary matrix $P$ that $P^2=1$ and $P\hat D^\prime P^{-1}=\hat D$. Then from $\hat D\,P\bar\Psi^\prime(t,r,\theta^\prime,\varphi^\prime)=P\hat D^\prime P^{-1}\,P\bar \Psi^\prime(t,r,\theta^\prime,\varphi^\prime)=0$ it follows that $P\bar\Psi^\prime(t,r,\theta^\prime,\varphi^\prime)=\bar\Psi(t,r,\theta,\varphi)$, or $\bar\Psi^\prime(t,r,\theta,\varphi)=P\,\bar\Psi(t,r,\pi-\theta,\pi+\varphi)$. From a comparison of (\ref{eq29}) and (\ref{eq30}) it is clear that $P$ must commute with $\gamma^0$, $\gamma^1$, and $\gamma^3$, and anticommute with $\gamma^2$. Such a $4\times 4$ matrix with a unit square has the form
\begin{equation}
 P=i\gamma^0\gamma^1\gamma^3=\left(\begin{array}{cc} 0 & -\sigma_1 \\ -\sigma_1 & 0 \end{array}\right )= \left(\begin{array}{cccc} 0 & 0 & 0 &-1 \\ 0 & 0 & -1 & 0 \\ 0 & -1 & 0& 0 \\ -1 & 0 & 0 & 0 \end{array}\right ).
 \label{eq31}
\end{equation}
Therefore, since $D^j_{-m,\,\pm\frac{1}{2}}(\pi+\varphi,\pi-\theta,0)=(-1)^j\,D^j_{-m,\,\mp\frac{1}{2}}(\varphi,\theta,0)$ \cite{Varshalovich_1988}, then the condition $\Psi^\prime(t,r,\pi-\theta,\pi+\varphi)=\delta_P\,(-1)^{j+1}\Psi(t,r,\theta,\varphi)$, $\delta_P=\pm 1$ requires \cite{Redkov_1998,Ovsiyuk_2011}
\begin{equation}
f_4=\delta_Pf_1,\;\; f_3=\delta_Pf_2.
\label{eq32}
\end{equation}
Introducing new functions \cite{Redkov_1998,Ovsiyuk_2011}
\begin{equation}
 \tilde f(t,r)=\frac{f_1(t,r)+f_2(t,r)}{\sqrt{2}},\;\;\;  \tilde g(t,r)=\frac{f_1(t,r)-f_2(t,r)}{i\sqrt{2}},
\label{eq33}
\end{equation}
we finally obtain from (\ref{eq28}) and (\ref{eq32})
\begin{eqnarray} &&
\left (\frac{d}{dr}+\frac{j+\frac{1}{2}}{S_\kappa(r)}\right )\tilde f+\left [i\left(\frac{\partial}{\partial t}+ieA_0\right)+\delta_P\,m\right]\tilde g=0, \nonumber \\ &&
\left (\frac{d}{dr}-\frac{j+\frac{1}{2}}{S_\kappa(r)}\right )\tilde g-\left [i\left(\frac{\partial}{\partial t}+ieA_0\right)-\delta_P\,m\right]\tilde f=0.
\label{eq34}
\end{eqnarray}

\section{Pauli equation in spaces of constant curvature}
For stationary states $\tilde f(t,r)=e^{-i(m+\epsilon)t}\,f(r)$ and $\tilde g(t,r)=e^{-i(m+\epsilon)t}\,g(r)$, where $m$ is the rest energy and $\epsilon$ is binding energy. The Coulomb potential for hydrogen-like atoms in spaces of constant curvature is ($e<0$ is the electron charge) \cite{Alizzi_2025}
\begin{equation}
    A_0=\frac{-e}{T_\kappa(r)}.
    \label{eq35}
\end{equation}
Therefore, (\ref{eq34}) takes the form
\begin{eqnarray} &&
\left (\frac{d}{dr}+\frac{j+\frac{1}{2}}{S_\kappa(r)}\right )f+\left [m+\epsilon+\frac{e^2}{T_\kappa(r)}+\delta_P\,m\right]g=0, \nonumber \\ &&
\left (\frac{d}{dr}-\frac{j+\frac{1}{2}}{S_\kappa(r)}\right )g-\left [m+\epsilon+\frac{e^2}{T_\kappa(r)}-\delta_P\,m\right] f=0.
\label{eq36}
\end{eqnarray}
In the non-relativistic approximation
\begin{equation}
\epsilon+\frac{e^2}{T_\kappa(r)}\ll m,
\label{eq37}
\end{equation}
equations (\ref{eq36}) show that if $\delta_P=1$, then $f\gg g$, and if $\delta_P=-1$, then $g\gg f$. 
Therefore, if $\delta_P=1$, up to linear order in small quantities, (\ref{eq36}) takes the form
\begin{eqnarray} &&
\left (\frac{d}{dr}+\frac{j+\frac{1}{2}}{S_\kappa(r)}\right )f+2m\,g=0, \nonumber \\ &&
\left (\frac{d}{dr}-\frac{j+\frac{1}{2}}{S_\kappa(r)}\right )g-\left [\epsilon+\frac{e^2}{T_\kappa(r)}\right] f=0.
\label{eq36a}
\end{eqnarray}
From the first equation we can express $g$ in terms of $f$ and its derivative, and substitute in the second equation.
Since
\begin{equation}
 \left (\frac{d}{dr}\mp\frac{\nu}{S_\kappa(r)}\right ) \left (\frac{d}{dr}\pm\frac{\nu}{S_\kappa(r)}\right ) =\frac{d^2}{dr^2}-\frac{\nu[\,\nu\pm C_\kappa(r)\,]}{S^2_\kappa(r)},\;\;\;\nu=j+\frac{1}{2},
 \label{eq38}
\end{equation}
then in the non-relativistic approximation from (\ref{eq36}) we obtain the radial Pauli equation for the large component $f$:
\begin{equation}
\frac{d^2 f}{dr^2}-\left (\frac{\nu\,[\,\nu+C_\kappa(r)\,]}{S^2_\kappa(r)}-2m\epsilon-\frac{2m e^2}{T_\kappa(r)}\right )f=0.
\label{eq39}
\end{equation}
Since $f_1=f_2\approx f/\sqrt{2}$, the correspondent two-component Pauli spinor is \cite{Ovsiyuk_2011}
\begin{equation}
 \Psi_{jm,\,\delta_P=1}(t,r,\theta,\varphi)=\frac{f(r)\,e^{-i(m+\epsilon)t}}{\sqrt{2}\,S_\kappa(r)}\left (\begin{array}{c} D_{-m,-1/2}^j(\varphi,\theta,0) \\ D_{-m,\;1/2}^j(\varphi,\theta,0) \\ \end{array}\right ),
 \label{eq40}
\end{equation}   
Similarly, if $\delta_P=-1$, then the large component is $g$, $f_3=-f_2\approx i\,g/\sqrt{2}$, the correspondent two-component Pauli spinor is 
\begin{equation}
 \Psi_{jm,\,\delta_P=-1}(t,r,\theta,\varphi)=\frac{i\,g(r)\,e^{-i(m+\epsilon)t}}{\sqrt{2}\,S_\kappa(r)}\left (\begin{array}{c} -D_{-m,\;1/2}^j(\varphi,\theta,0) \\ \;\;D_{-m,-1/2}^j(\varphi,\theta,0) \\ \end{array}\right ),
 \label{eq41}
\end{equation}
and the equation for $g$ looks like
\begin{equation}
\frac{d^2 g}{dr^2}-\left (\frac{\nu\,[\,\nu-C_\kappa(r)\,]}{S^2_\kappa(r)}-2m\epsilon-\frac{2m e^2}{T_\kappa(r)}\right )g=0.
\label{eq42}
\end{equation}
Since this equation can be obtained from (\ref{eq39}) by replacing $\nu\to -\nu$, $f\to g$, it will be sufficient to consider only (\ref{eq39}). 

It is convenient to introduce dimensionless variables (we assume $\kappa\ne 0$)
\begin{equation}
    z=e^{i\sqrt{\kappa}\,r},\;\;\bar\epsilon=\frac{2m\epsilon}{\kappa},\;\;\bar\lambda=\frac{2ime^2}{\sqrt{\kappa}},\;\;\bar\nu=2\nu=2j+1.
\label{eq43}    
\end{equation}
Then equation (\ref{eq39}) can be rewritten as follows
\begin{equation}
 \frac{d^2f}{dz^2}+\frac{\pi_1(z)}{\sigma(z)}\,\frac{df}{dz}+\frac{\sigma_1(z)}{\sigma^2(z)}\,f=0,
 \label{eq44}
\end{equation}
where
\begin{eqnarray} &&
\sigma(z)=z(1-z^2),\;\pi_1(z)=1-z^2, \nonumber \\ &&
\sigma_1(z)=\bar\lambda-\bar\epsilon-\bar\nu z+(2\bar\epsilon-\bar\nu^2)z^2-\bar\nu z^3-(\bar\lambda+\bar\epsilon)z^4.  
\label{eq45}
\end{eqnarray}
As can be seen from its form, equation (\ref{eq44}) is of generalized Heun type and can be investigated by the extended Nikiforov-Uvarov method \cite{Karayer_2015,Quesne_2017}. The standard Nikiforov-Uvarov method is not applicable to equation (\ref{eq44}), since it assumes that $\sigma(z)$ and $\sigma_1(z)$ are polynomials of degree no higher than two, and $\pi_1(z)$ is a polynomial of degree no higher than one, whereas in our case $\sigma(z)$ is a polynomial of degree three, $\sigma_1(z)$ is a polynomial of degree four, and $\pi_1(z)$ is a polynomial of degree two.

\section{Pauli equation and extended  Nikiforov-Uvarov method}
The extended Nikiforov-Uvarov method is described in \cite{Karayer_2015,Quesne_2017}. In general, it resembles the well-known algorithm of the standard Nikiforov-Uvarov method, described in detail, for example, in \cite{Alizzi_2025}. We illustrate the application of the extended Nikiforov-Uvarov method to the equation (\ref{eq44}). The set of solutions to equation (\ref{eq44}) is invariant under a ``gauge'' transformation
\begin{equation}
f(z)=e^{\phi(z)}y(z),    
\label{eq46}
\end{equation}
provided
\begin{equation}
\frac{d\Phi}{dz}\equiv\Phi^\prime=\frac{\pi(z)}{\sigma(z)},
\label{eq47}
\end{equation}
where $\pi(z)$ is some polynomial, maximum of the second degree. In this case, the function $y(z)$  also satisfies the equation of generalized Heun type
\begin{equation}
y^{\prime\prime}+\frac{\tau(z)}{\sigma(z)}\,y^\prime+\frac{\sigma_2(z)}{\sigma^2(z)}\,y=0,
\label{eq48}    
\end{equation}
where
\begin{equation}
\tau(z)=\pi_1(z)+2\pi(z)
\label{eq49}
\end{equation}
is a polynomial of degree no higher than two, and
\begin{equation}
\sigma_2(z)=\sigma_1(z)+\pi^2(z)+\pi(z)\left [\pi_1(z)-\sigma^\prime(z)\right]+\pi^\prime(z)\sigma(z)
\label{eq50}
\end{equation}
is a polynomial of degree no higher than four.

By using the freedom to choose the polynomial $\pi(z)$, equation (\ref{eq48}) can be simplified. In particular, if $\pi(z)$ is chosen such that
\begin{equation}
\frac{\sigma_2(z)}{\sigma(z)}=h(z),
\label{eq51}
\end{equation}
where $h(z)$ is a polynomial of degree at most one, equation (\ref{eq48}) reduces to the following form:
\begin{equation}
\sigma(z)\,y^{\prime\prime}+\tau(z)\,y^\prime+h(z)\, y=0.
\label{eq52}
\end{equation}
Equations (\ref{eq50}) and (\ref{eq51}) imply that $\pi(z)$ is the root of a quadratic equation whose solution is
\begin{equation}
\pi(z)=\frac{\sigma^\prime(z)-\pi_1(z)}{2}\pm\sqrt{\left(\frac{\sigma^\prime(z)-\pi_1(z)}{2}\right)^2-\sigma_1(z)+g(z)\sigma(z)}\,,
\label{eq53}
\end{equation}
where $g(z)=h(z)-\pi^\prime(z)$ is a polynomial of degree at most one. Only if
\begin{equation}
\sigma_3(z)=\left(\frac{\sigma^\prime(z)-\pi_1(z)}{2}\right)^2-\sigma_1(z)+g(z)\,\sigma(z) \label{eq54}
\end{equation}
is the square of a polynomial of at most second-degree, then $\pi(z)$ will be a polynomial of at most second-degree. In our case, taking $g(z)=g_0+g_1z$, where $g_0,\,g_1$ are some constants, we get
\begin{equation}
\pi(z)=-z^2\pm\sqrt{\sigma_3(z)},
\label{eq55}
\end{equation}
where
\begin{equation}
\sigma_3(z)=(1+\bar\lambda+\bar\epsilon-g_1)z^4+(\bar \nu-g_0)z^3+(\bar\nu^2-2\bar\epsilon+g_1)z^2+(\bar \nu+g_0)z+\bar\epsilon-\bar\lambda.
\label{eq56}
\end{equation}
We will have $\sigma_3(z)=(az^2+bz+c)^2$ if the coefficients $a,\,b,\,c$ satisfy the relations
\begin{eqnarray} &&
a^2=1+\bar\lambda+\bar\epsilon-g_1,\;b^2+2ac=\bar\nu^2-2\bar\epsilon+g_1,
\nonumber \\ && 2ab=\bar\nu-g_0,\;\;\; 2bc=\bar\nu+g_0,\;\;\;c^2=\bar{\epsilon}-\bar{\lambda}.   
\label{eq57}
\end{eqnarray} 
It follows from these equations that
\begin{equation}
(a+c)^2+b^2=1+\bar\nu^2,\;\;\;(a+c)b=\bar\nu.
\label{eq58}
\end{equation}
Together with the last equation in (\ref{eq57}), the relations (\ref{eq58}) completely determine the coefficients $a$, $b$, and $c$ (but not uniquely). In particular, (\ref{eq58}) indicates that $(a+c)^2$ and $b^2$ are roots of the quadratic equation $$x^2-(1+\bar{\nu}^2)x+\bar{\nu}^2=0,$$ and since according to the second relation in (\ref{eq58}) $a+c$ and $b$ have the same sign, we are left with the following four possibilities:
\begin{eqnarray} &&
(1)\;\;a+c=\bar{\nu},\;b=1; \;\;\;\;(2)\;\; a+c=-\bar{\nu},\;b=-1;
\nonumber \\ &&
(3)\;\;a+c=1,\;b=\bar{\nu};\;\;\;\;(4)\;\; a+c=-1,\;b=-\bar{\nu}.
\label{eq58a}
\end{eqnarray}
Therefore, $\pi(z)=(-1\pm a)z^2\pm bz\pm c$ and equation (\ref{eq47}) takes the form
\begin{equation}
\frac{d\Phi}{dz}=\frac{\pi(z)}{\sigma(z)}=\frac{A}{z}+\frac{B}{z-1}+\frac{C}{z+1},
\label{eq59}
\end{equation}
where
\begin{equation}
\;A=\pm c,\;\;B=\frac{1\mp (a+b+c)}{2},\;\;C=\frac{1\mp (a+c-b)}{2}.
\label{eq60}
\end{equation}
Then from (\ref{eq46})
\begin{equation}
f(z)=z^A\,(z-1)^B\,(z+1)^C\,y(z),
\label{eq61} 
\end{equation}
and to avoid singularities at $z=\pm 1$, we choose positive $B$, and $C$. In particular, the inspection of (\ref{eq58a}) shows that only first two options can be used for this goal without severely restricting possible values of $\bar{\nu}$. First
we take $a+c+b=1+\bar\nu$, $a+c-b=\bar\nu-1$, and the lower ``+'' signs in (\ref{eq60}), which leads to
\begin{eqnarray} &&
c=-\sqrt{\bar\epsilon -\bar\lambda},\;\;b=1,\;\;a=\bar\nu-c,\nonumber \\ &&
A=-c,\;\;B=1+\frac{\bar\nu}{2}=j+\frac{3}{2},\;\;C=\frac{\bar\nu}{2}=j+\frac{1}{2}.
\label{eq62}
\end{eqnarray}
In the hyperbolic case ($\kappa<0)$, $z\to 0$ as $r\to\infty$. Therefore, $A$ must also be positive to ensure acceptable behavior near $z=0$.
 
The Pauli equation in the form (\ref{eq52}) can be rewritten as the standard Heun equation \cite{Slavyanov_2000,Smirnov_2002,Karayer_2015}
\begin{equation}
y^{\prime\prime}+\left [\frac{\gamma}{z}+\frac{\delta}{z-1}+\frac{\varepsilon}{z+1}\right ]\,y^\prime+\frac{\alpha\beta\,z-q}{z(z-1)(z+1)}\,y=0,
\label{eq63}
\end{equation}
where $\gamma$, $\delta$, $\varepsilon$, $\alpha$ and $\beta$ are some constants satisfying the Fuchsian condition $\varepsilon=\alpha+\beta-\gamma-\delta+1$, which ensures the regularity of the singularity at infinity. Indeed, taking $\pi(z)=-(1+a)z^2-z-c$ and writing $h(z)=-\alpha\beta\,z+q=g_0-1+(g_1-2-2a)z$, on the one hand, in light of (\ref{eq59}) we will have:
\begin{equation}
\frac{\tau(z)}{\sigma(z)}=\frac{\pi_1(z)}{\sigma(z)}+2\frac{\pi(z)}{\sigma(z)}=\frac{2A+1}{z}+\frac{2B}{z-1}+\frac{2C}{z+1},
\label{eq64}
\end{equation}
and on the other hand, in light of (\ref{eq57}) and (\ref{eq62}),
\begin{equation}
q=g_0-1=b(c-a)-1=-1-\bar{\nu}-2\sqrt{\bar{\epsilon}-\bar{\lambda}},
\label{eq65}
\end{equation}
and
\begin{equation}
\alpha\beta=2(1+a)-g_1=(1+a)^2-(\bar{\epsilon}+\bar{\lambda}),
\label{eq66}
\end{equation}
while the Fuchsian condition will give
\begin{equation}
    \alpha+\beta=2(A+B+C)=2(-c+1+\bar{\nu})=2(1+a).
\label{eq67}    
\end{equation}
Equations (\ref{eq66}) and (\ref{eq67}) show that $\alpha$ and $\beta$ are solutions of the quadratic equation $x^2-2(1+a)x+(1+a)^2-(\bar{\epsilon}+\bar{\lambda})$. Therefore, we can take
\begin{eqnarray} &&
\alpha=1+a-\sqrt{\bar{\epsilon}+\bar{\lambda}}=1+\bar{\nu}+\sqrt{\bar{\epsilon}-\bar{\lambda}}-\sqrt{\bar{\epsilon}+\bar{\lambda}}, \nonumber \\ &&   
\beta=1+a+\sqrt{\bar{\epsilon}+\bar{\lambda}}=1+\bar{\nu}+\sqrt{\bar{\epsilon}-\bar{\lambda}}+\sqrt{\bar{\epsilon}+\bar{\lambda}}.
\label{eq68}
\end{eqnarray}
Expressions (\ref{eq65}) and (\ref{eq68}) are consistent with the results of \cite{Ovsiyuk_2011}.

Polynomiality is not strictly required by normalizability or single-valuedness of the wave function. However, in the context of extended Nikiforov-Uvarov method, we are interested in polynomial solutions of (\ref{eq52}) that exist under a certain quantization condition. This quantization condition can be obtained in the following way \cite{Karayer_2015}. 
The derivatives $v_n(z)=y^{(n)}(z)$ of the function $y(z)$ satisfy the following third-order differential equation
\begin{equation}
 \sigma(z)v_n^{\prime\prime\prime}+\mu_n(z)v_n^{\prime\prime}+\tau_n(z) v_n^{\prime}+h_n(z) v_n=0,
 \label{eq69}
\end{equation}
where
\begin{eqnarray} &&
\mu_n(z)=\tau(z)+(n+1)\sigma^\prime(z), \nonumber \\ &&
\tau_n(z)=(n+1)\tau^\prime(z)+\frac{1}{2}n(n+1)\sigma^{\prime\prime}(z)+h(z),  \\ &&
h_n(z)=(n+1)h^\prime(z)+\frac{1}{2}n(n+1)\tau^{\prime\prime}(z)+\frac{1}{6}n(n+1)(n-1)\sigma^{\prime\prime\prime}(z). \nonumber
\label{eq70}
\end{eqnarray}
The relation (\ref{eq69}) can be proved by induction. Indeed, differentiating (\ref{eq52}), we obtain (\ref{eq69}) for $n=0$. On the other hand, differentiation of (\ref{eq69} leads to the following recurrence relations
\begin{equation}
\mu_{n+1}(z)=\mu_n(z)+\sigma^\prime(z),\; \tau_{n+1}(z)=\tau_n(z)+\mu^\prime(z),\;  h_{n+1}(z)=h_n(z)+\tau^\prime(z),
\label{eq71}
\end{equation}
and if (\ref{eq70}) is true for $n$, then one can easily verify that (\ref{eq71}) ensures that these relations are also true for $n+1$.

If $y(z)$ is a polynomial of degree $n$, then $v_n(z)=\mathrm{const}$ and equation (\ref{eq69}) can only be true if $h_n(z)=0$. This circumstance produces the desired quantization condition:
\begin{equation}
h^\prime(z)=-\frac{1}{2}n\,\tau^{\prime\prime}(z)-\frac{1}{6}n(n-1)\,\sigma^{\prime\prime\prime}(z).    
\label{eq72}
\end{equation}
In our case 
\begin{equation}
\pi(z)=-(1+a)z^2-z-c,\;\;\;\tau(z)=1-2c-2z-(3+2a)z^2,
\label{eq73}
\end{equation}
and (\ref{eq72}) reduces to $\alpha\beta=-n(2+2a+n)$. In light of (\ref{eq66}), this last equation implies $\bar\epsilon+\bar\lambda=(1+a+n)^2$. Therefore, using $a=\bar\nu+\sqrt{\bar{\epsilon}-\bar\lambda}$ from (\ref{eq62}), we obtain
\begin{equation}
2(1+n+\bar\nu)\sqrt{\bar \epsilon-\bar\lambda}=2\bar\lambda-(1+n+\bar\nu)^2,
\label{eq74}
\end{equation}
and, introducing the quantum number $N=1+n+\bar{\nu}=2(j+1)+n$,
\begin{equation}
\bar\epsilon=\frac{\bar{\lambda}^2}{N^2}+\frac{N^2}{4}.
\label{eq75}
\end{equation}
Remembering definition of $\bar\epsilon$ and $\bar\lambda$ (equation (\ref{eq43})), we get according to (\ref{eq75}) the following energy spectrum
\begin{equation}
\epsilon_N=Ry\left [-\frac{4}{N^2}+\frac{N^2}{4}\,\kappa \,a_B^2\right],\;\;Ry=\frac{1}{2ma_B^2},\;\; a_B=\frac{1}{me^2}.   
\label{eq76}
\end{equation}
However, not every natural number $N$ is admissible in (\ref{eq76}), since the wave function as a function of $r$ must be single-valued. In particular, in light of (\ref{eq43}) and (\ref{eq74}), equation (\ref{eq61}) implies that 
\begin{equation}
f(z)=e^{-\frac{2}{Na_B}}\,z^{-\frac{N}{2}}\,(z-1)^{j+\frac{3}{2}}\,(z+1)^{j+\frac{1}{2}}\,y_n(z),
\label{eq77}
\end{equation}
and since $j=l\pm\frac{1}{2}$, this is a single-valued function of $z$ only if $N=2(l+1)+n\pm 1=2\bar{n}$ is an even number, that is, if $n$ is odd. The situation is entirely analogous to how half-integer values of the orbital angular momentum $l$ must be excluded as eigenvalues of $\hat{\bf{L}}^2$ for a spinless particle \cite{Merzbacher_1962,Pauli_1939}.

Therefore, $\bar{n}$ plays the role of the principal quantum number, and instead of (\ref{eq76}) we obtain
\begin{equation}
 \epsilon_{\bar{n}}=Ry\left [-\frac{1}{\bar{n}^2}+\bar{n}^2\,\kappa a_B^2\right],
\label{eq78}   
\end{equation}
a result that agrees well with the result obtained using the Schr\"{o}dinger equation \cite{Alizzi_2025}, with only one important exception: the ``geometric potential'' $-\frac{\kappa}{2m}$ is missing in (\ref{eq78}). This is a manifestation of a result well known for the Dirac equation on a curved surface: the non-relativistic limit does not commute with the thin-layer quantization method, or, equivalently, the non-relativistic limit does not commute with the ``squaring'' of the Dirac equation on a curved surface \cite{Brandt_2016,Wang_2022}.

The results obtained so far using the extended Nikiforov-Uvarov method appear reasonable and interesting. However, the following observation undermines the validity of this method. In contrast to ordinary Nikiforov-Uvarov method, the quantization condition (\ref{eq72}) provides only necessary, but not sufficient condition for the differential equation (\ref{eq52}) to have polynomial solutions. Indeed, assuming $y(z)=\sum\limits_{m=0}^n C_m z^m$ and $\tau(z)$ from (\ref{eq73}) in (\ref{eq52}), we obtain the following system of equations for polynomial coefficients $C_m$:
\begin{eqnarray} &&
(m+1)(m+1-2c)C_{m+1}-[(m-1)(1+2a+m)+\alpha\beta]C_{m-1}+
\nonumber \\ &&
(q-2m)C_m=0,\;\;0\le m\le n, \,\,\,\,\,\, C_{-1}=C_{n+1}=0,
\nonumber \\ &&
\alpha\beta=-n(n+2+2a).
\label{eq80}
\end{eqnarray}
The quantization condition (\ref{eq72}) gives only the last equation. But to have the non-zero solution of the system  (\ref{eq80}) the determinant of the corresponding tri-diagonal matrices
%\begin{equation}
%\mbox{\fontsize{4pt}{5pt}\selectfont
%\colorbox{yellow}{$\left |\begin{array}{ccccccccc} q & 1-2c & 0 & 0 & 0 & . & . & . & 0\\-\alpha\beta & q-2 & 4(1-c) & 0 & 0 & . & . & . & 0\\0 & -[\alpha\beta+3+2a] & q-4 & 3(3-2c) & 0 & . & . & . & 0\\ . & %.& . & . & . & . & . & .& .\\  . & .& . & . & . & . & . & .& . \\ 0 & 0 & 0 & . & . & 0 & -[\alpha\beta+(n-2)(n+2a)] & q-2n+2 & n(n-2c) \\ 0 & 0 & 0 & . & . & 0 & 0 & -[\alpha\beta+(n-1)(1+2a+n)] & q=2n
%\end{array}\right |$}}    
%\label{eq81}
%\end{equation}
\begin{equation}
\begin{pmatrix}
  b_0 & c_1 & & \\
  a_1 & b_1 & \ddots & \\
  & \ddots & \ddots & c_n \\
  & & a_n & b_n
\end{pmatrix},
\label{eq81}
\end{equation}
where
\begin{eqnarray} &&
a_m=-[\alpha\beta+(m-1)(m+1+2a)],\;m=1,2,\ldots,n,
\nonumber \\ &&
b_m=q-4m,\; m=0,1,\ldots,n,
\nonumber \\ &&
c_m=m(m-2c),\;m=1,2,\ldots,n,
\label{eq82}
\end{eqnarray}
must be zero. This condition restricts the accessory parameter $q$ to some specific values for which polynomial solutions of the Heun equation exist, and this is characteristic to  quasi-exactly solvable spectral problems, which are  particular cases of Heun’s equation \cite{Olver_1994,Ushveridze_1994}.

In our case the accessory parameter is already restricted by the quantization condition (\ref{eq72}), and the second, zero-determinant condition, cannot be satisfied. Indeed, for $n=1$ this latter condition is
\begin{equation}
\Delta_1=\left |\begin{array}{cc} q & 1-2c \\-\alpha\beta & q-2 \end{array}\right |=q(q-2)+\alpha\beta(1-2c)=0.    
\label{eq83}   
\end{equation}
But from (\ref{eq65}) $q=c-a-1$, and from quantization condition, for $n=1$, $\alpha\beta=-(3+2a)$. Substituting this into $\Delta_1$ will give 
$$\Delta_1=(c+a)(c+a+2)=\bar{\nu}(\bar{\nu}+2)\ne 0.$$ 
Therefore, unfortunately, we do not have a solid support for the nice spectrum (\ref{eq78}), since it was obtained under supposition that (\ref{eq52}) has a polynomial solution.

It can be verified that if we choose $a+c=-\bar\nu$, $b=-1$, the upper ``-'' signs in (\ref{eq60}), and $A=c=\sqrt{\bar\epsilon -\bar\lambda}$, the quantization condition leads to the same spectrum (\ref{eq78}), but with the same attendant problems with polynomial solutions.

\section{Concluding remarks}
The Heun differential equation, which is the most general second order differential equation with four regular singular points, finds application in a number of physical problems \cite{Hortascu_2018}, as diverse as celestial mechanics, quantum mechanics, quantum field theory, atomic and nuclear physics, gravitational physics, black holes, astrophysics, cosmology and many others \cite{Fiziev_2015,Levai_2025,Chen_20211,Berkdemir_2012,Slavyanov_2000}. The useful and elegant Nikiforov-Uvarov method, well known in the theory of hypergeometric type equations, was extended to Heun type equations in \cite{Karayer_2015} and applied to some quantum mechanical problems (see, for example, \cite{Ikot_2022,Karayer_2018,Likene_2023,Karayer_2022}). A nonrelativistic particle with spin 1/2, bound by a Coulomb potential in spaces of constant curvature, is also described by the Heun equation \cite{Ovsiyuk_2011}. At first glance, it seems that the extended Nikiforov-Uvarov method is applicable in this case and leads to a significant simplification. Comparison of the obtained energy spectrum with the spectrum from \cite{Alizzi_2025} demonstrates the non-commutativity of the non-relativistic limit with the ``squaring'' of the Dirac equation in the case of curved spatial geometry. Unfortunately, due to the peculiarities of the extended Nikiforov-Uvarov method, these result lacks mathematical rigor.

Although the problem of quantization on spacelike curved surfaces is of fundamental importance, helping to understand the introduction of the Hamiltonian formalism with its special role of time in relativistic situations \cite{Dirac_2001}, the problem discussed in this article may seem purely academic. However, due to the rapid development of technologies for creating artificial micro- and nanostructures, quantum mechanics, both nonrelativistic and relativistic, on curved surfaces is becoming an active area of research \cite{Wang_2022,Brandt_2016}.

It should be noted that the application of the extended Nikiforov-Uvarov method to some quasi-exactly solvable potentials in \cite{Karayer_2018} was criticized in \cite{Fernandez_2026} with the rather radical conclusion that the extended Nikiforov-Uvarov method is “unsuitable for solving eigenvalue equations” (see also \cite{Fernandez_2021}). In our case, at first glance, the extended Nikiforov-Uvarov method correctly reproduces the results obtained by the more traditional method in \cite{Ovsiyuk_2011}, and simplifies their derivation. So it would seem that this method would be useful, at least in some circumstances. However, in the context of the Nikiforov-Uvarov method, there is a fundamental difference between hypergometric-type equations and Heun-type equations. In the traditional Nikiforov-Uvarov method, the quantization condition inherent in this method is sufficient to ensure the existence of polynomial solutions, whereas for Heun-type equations, the quantization condition of the extended Nikiforov-Uvarov method provides only a necessary condition for the existence of polynomial solutions. Failure to appreciate this difference can lead to serious errors \cite{Fernandez_2026}.
Ultimately, the extended Nikiforov-Uvarov method is a technical tool, albeit a rather imperfect one, and like all technical tools, it should be used with due caution.

\section*{Acknowledgments}
We thank the anonymous reviewers for their comments, which helped improve the manuscript and change its focus.

\bibliographystyle{elsarticle-num-names} 
\bibliography{Pauli_CC.bib}

\end{document}